\documentclass[12pt]{elsart}
\usepackage{graphicx}
\usepackage{amssymb}
\usepackage{times}
\begin{document}

\newcommand{\re}{\mathop{\mathrm{Re}}}
\newcommand{\im}{\mathop{\mathrm{Im}}}
\newcommand{\D}{\mathop{\mathrm{d}}}
\newcommand{\I}{\mathop{\mathrm{i}}}

\def\lambar{\lambda \hspace*{-5pt}{\rule [5pt]{4pt}{0.3pt}} \hspace*{1pt}}

\noindent {\Large DESY 10-011}

\noindent {\Large January 2010}

\bigskip

\begin{frontmatter}

\date{}

\title{
Possible operation of the European XFEL with ultra-low emittance beams}

\author{R.~Brinkmann,}
\author{E.A.~Schneidmiller}
\author{and M.V.~Yurkov}

\address{Deutsches Elektronen-Synchrotron (DESY),
Notkestrasse 85, D-22607 Hamburg, Germany}

\begin{abstract}
Recent successful lasing of the Linac Coherent Light Source (LCLS) in the hard x-ray
regime and the experimental demonstration of a possibility to produce low-charge
bunches
with ultra-small normalized emittance have lead to the discussions on optimistic scenarios of
operation of the European XFEL.
In this paper we consider new options that make use of low-emittance beams, a relatively high
beam energy, tunable-gap undulators, and a multi-bunch capability of this facility.
We study the possibility of operation
of a spontaneous radiator (combining two of them, U1 and U2, in one beamline) in the SASE mode in
the designed photon energy range 20-90 keV and show that it becomes
possible with ultra-low emittance electron beams similar to those generated in LCLS.
As an additional attractive option we consider
the generation of powerful soft x-ray and VUV radiation by the same electron bunch
for pump-probe experiments, making use of recently invented compact afterburner scheme.
We also propose a betatron switcher as a simple, cheap, and robust solution
for multi-color operation of SASE1 and SASE2
undulators, allowing to generate 2 to 5 x-ray beams of different independent colors from each of
these undulators for simultaneous multi-user operation. We describe a scheme for
pump-probe experiments, based on a
production of two different colors by two closely spaced electron bunches (produced in
photoinjector) with the help of a very fast betatron switcher.
Finally, we discuss how without significant modifications of the layout the European XFEL can
become a unique facility that continuously covers with powerful, coherent
radiation a part of the electromagnetic spectrum from far infrared to gamma-rays.
\end{abstract}

\end{frontmatter}

\baselineskip 20pt

\clearpage

\section{Introduction}
Free-electron lasing at wavelengths shorter than the ultraviolet
can be achieved with a single-pass, high-gain
FEL amplifier operating in the so called Self-Amplified Spontaneous
Emission (SASE) mode, where the amplification process starts from
shot noise in the electron beam \cite{kondratenko-1980,derbenev-xray-1982,pellegrini}. Present
accelerator and FEL techniques allow to
generate powerful, coherent femtosecond pulses in the wavelength range from vacuum
ultraviolet (VUV) \cite{ttf1,32nm,spring8} through soft x-ray \cite{photonics,njp}
to hard x-ray \cite{emma-epac}.

Recent successful lasing of the Linac Coherent Light Source (LCLS) in the hard x-ray
regime \cite{emma-epac} and the demonstration of the possibility to produce low-charge bunches (20 pC)
with ultra-small normalized slice emittance (0.15 mm mrad) \cite{20pC} have lead to
the discussions on optimistic scenarios of operation of the European XFEL. In this paper we
study new options that make use of
low-emittance beams, a relatively high beam energy, and multi-bunch capability
of this facility.

The European XFEL \cite{euro-xfel-tdr} will serve several users simultaneously with x-ray
radiation produced in different undulators. In particular, two undulators (U1 and U2, both having
50 m net magnetic length) are supposed to use
spent electron beam from SASE2 undulator
and to produce spontaneous radiation in the photon energy range 20-90 keV \cite{euro-xfel-tdr},
which is of high interest for materials science studies.
In Section 2 we consider
SASE process in a "spontaneous radiator", called below SASE-U1. Of course, the
bunches should not be disturbed by FEL interaction in SASE2 undulator,
so that one has to use a fast kicker in front of SASE2 to kick selected bunches by an angle that is
sufficient
for suppression of lasing to saturation of these bunches. Then these bunches acquire in SASE2 only
a relatively small increase of an energy spread due to quantum diffusion depending on operating
wavelength of SASE2. We consider an electron beam with the normalized emittance 0.15 mm mrad and
a peak
current 5 kA, and show that the saturated FEL power at the highest photon energy, 90 keV, can be
obtained at the undulator length of 70-100 m, depending on electron beam energy (we consider
standard operating energy 17.5 GeV, and a more preferable scenario with a higher energy, 22 GeV).
So, the operation
of SASE FEL at the highest design photon energy would require to combine two undulators (U1 and U2)
in one beamline. At 60 keV, for example, saturation can be reached within 50 m, so that even
one undulator might be sufficient. Saturated power is in the range of tens of gigawatts, and pulse
duration is few femtoseconds. We study the dependence of the highest available photon energy
on emittance and show that 50-70 keV can be reached if normalized slice
emittance is 0.3-0.4 mm mrad.

As an additional attractive option we consider the operation of a recently invented
afterburner \cite{ab-fel,ab-prst}, installed behind SASE-U1.
In Section 3 we show that for a total length of such an afterburner (including a chicane and
an undulator) of about 10 meters, one can produce intense, sub-gigawatt-level VUV (vacuum ultraviolet)
and soft x-ray pulses in the present wavelength range of FLASH \cite{njp}. Both soft and
hard x-ray pulses are perfectly synchronized, have femtosecond duration and can be used in a
pump-probe experiment, or can be separated and used by different experimental stations.

The lengths of SASE1 and SASE2 undulators were calculated \cite{euro-xfel-tdr}
under assumption of rather
conservative normalized emittance, 1.4 mm mrad.
In Section 4 we calculate saturation length in these undulators as a function of wavelength
and normalized emittance, assuming the latter parameter to be significantly reduced.
With bright electron beams the full length of SASE1 and SASE2 undulators can be efficiently used to
generate multiple x-ray beams
with different independent wavelengths for simultaneous multi-user operation. In Section 5
we propose a betatron switcher and show
that one only needs to install a compact (about 1 m long) fast
kicker\footnote{Using the fast intra-bunch
feedback system in feed-forward mode may be another option to realize
this scheme without having to install any additional hardware.} in front of a considered undulator
without any modifications of the undulator itself. Different groups of bunches
get different angular kicks, and
for every group a kick is compensated statically (by corrections coils or moving quadrupoles)
in a part of the undulator (sub-undulator),
tuned to the wavelength designated to the given group.
As a result, two to five colors can
be simultaneously generated from each undulator (depending on emittance and shortest desirable
wavelength) with pulse patterns requested by each of the independent users.
The scheme is very flexible and is operated remotely from the control room with a quick change of
a desired set of photon energies.

As a generalization of the method of the betatron switcher,
we briefly describe a scheme for pump-probe experiments, making use of
closely spaced electron bunches, produced in photoinjector. In front of the undulator they
get different angular kicks from a transverse deflecting cavity, and then lase in different
parts of the undulator.

Finally, in Section 6 we discuss how the European XFEL facility can cover with powerful, coherent
radiation a part of the electromagnetic spectrum from far infrared to gamma-rays.

\section{Parameters of SASE-U1}

\begin{figure*}[b]

\includegraphics[width=0.8\textwidth]{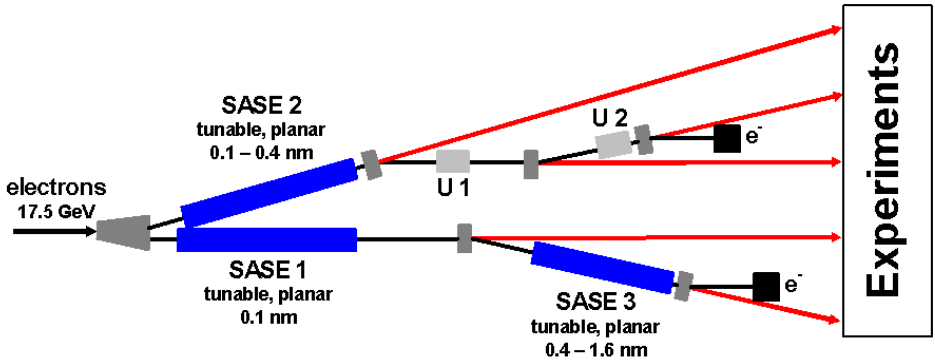}

\caption{\small Layout of the European XFEL.}

\label{xfel}
\end{figure*}

Two undulators (U1 and U2, both having
50 m net magnetic length) are supposed to use
spent electron beam from SASE2 undulator of the European XFEL (see Fig.~\ref{xfel})
and to produce spontaneous radiation in the photon energy range 20-90 keV \cite{euro-xfel-tdr}.
However, after
successful lasing of the Linac Coherent Light Source (LCLS) in the hard x-ray
regime \cite{emma-epac} and the experimental demonstration of a possibility to produce low-charge
bunches with ultra-small normalized slice emittance \cite{20pC}, one can think of a new approach
to this photon energy range. Here we consider SASE process in this range, assuming that both
undulators are combined in one beamline. The net magnetic length of this combined undulator
(which we call SASE-U1) is 100 m, the total length is 122 m. Parts of beamline for U1 and U2 are,
respectively, 190 m and 250 m long (see Table~\ref{tab:xfel-tunnels}),
so that one can find a place for such an undulator.

\begin{table}[b]
\caption{Location of undulators on XFEL site}
\bigskip

\begin{tabular}{ l l l l }
\hline
Location & Full length & Available & Undulator \\
	 &             &   for     & \\
	 &             & undulators    & \\
\hline \\
XS1-XS3   & 620 m & 396 m & SASE1 \\
XS3-XHDU1 & 301 m & 251 m & SASE3 \\
\hline \\
XS1-XS2   & 550 m & 358 m & SASE2 \\
XS2-XS4   & 190 m & & Spont. U1\\
XS4-XHDU2 & 250 m & & Spont. U2\\
\hline
\end{tabular}
\label{tab:xfel-tunnels}
\end{table}

\begin{figure*}[b]

\includegraphics[width=0.45\textwidth]{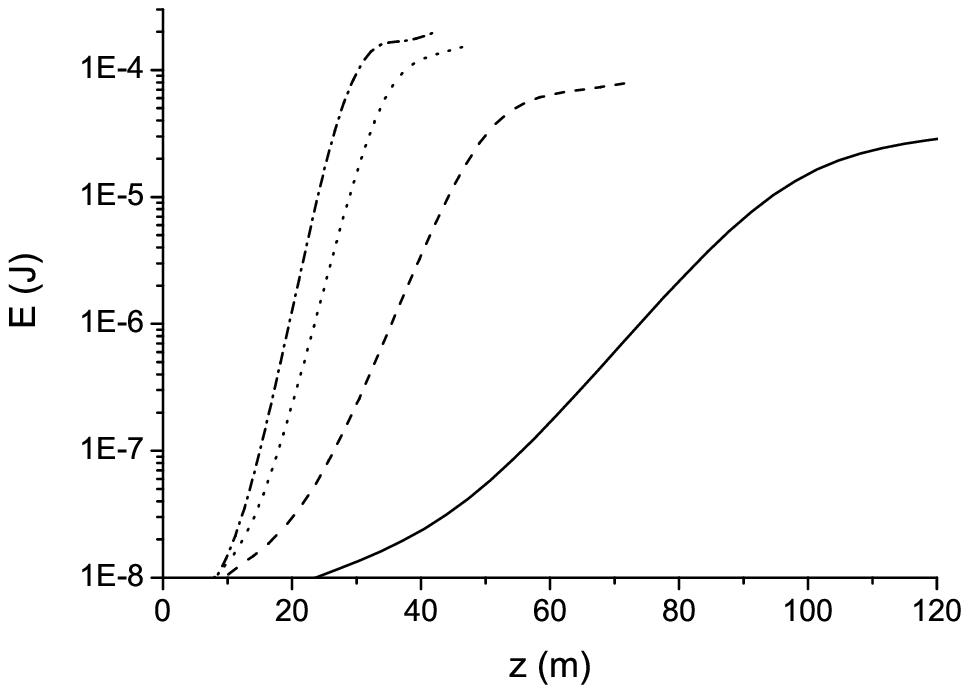}
\includegraphics[width=0.45\textwidth]{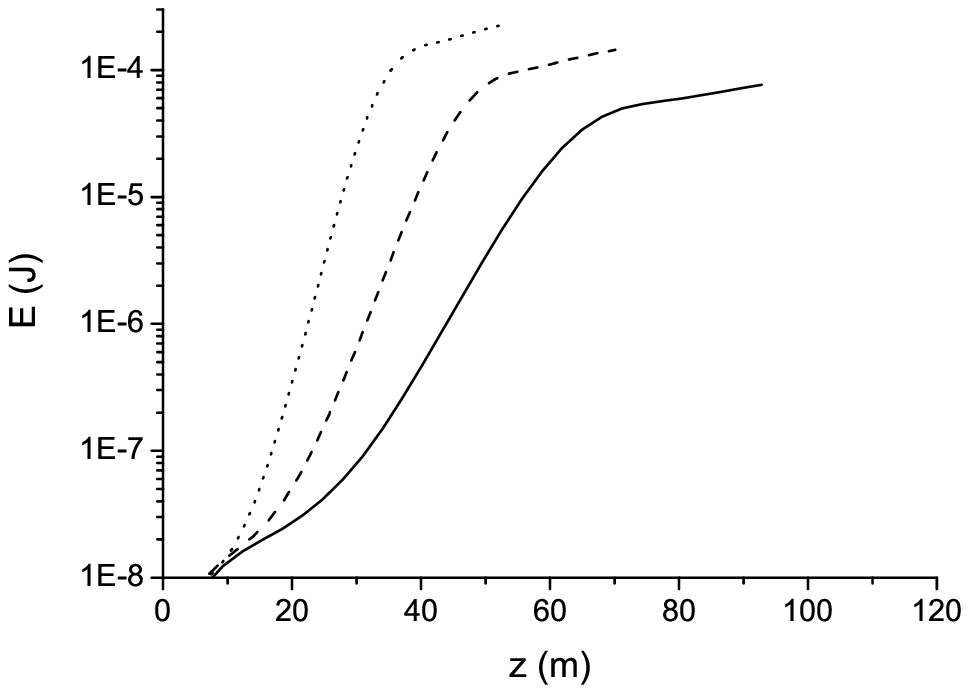}

\caption{\small Pulse energies versus net magnetic length of SASE-U1 undulator
for electron beam energies 17.5 GeV (left plot) and 22 GeV (right plot) and
different photon energies: 20 keV (dash-dot), 30 keV (dot), 60 keV (dash), and 90 keV (solid).}

\label{sat17-22}
\end{figure*}

\begin{table}[b]
\caption{Parameters of electron beam and SASE-U1 undulator}
\bigskip

\begin{tabular}{ l l }
\hline

Electron energy & 17.5 GeV / 22 GeV \\
Bunch charge    & 20 pC \\
Peak current    & 5 kA \\
Normalized slice emittance & 0.15 mm mrad \\
Slice energy spread & 1.7 MeV / 3 MeV \\
Beta-function & 15-25 m \\
Net undulator length & 100 m \\
Undulator period & 2.6 cm \\
Undulator K-parameter (rms) & 0.5-2.1 \\

\hline
\end{tabular}
\label{tab:sase-u1-par}
\end{table}

Of course, the
bunches should not be disturbed by the FEL interaction in the SASE2 undulator,
so that one has to use a fast kicker in front of SASE2 to kick selected bunches by an angle that is
sufficient
for suppression of lasing to saturation of these bunches. Then these bunches acquire in SASE2 only
a relatively small increase of an energy spread due to quantum diffusion \cite{q-diff}
depending on operating wavelength of SASE2. To be specific, in the following we assume that
SASE2 is tuned to 1 $\mathrm{\AA}$. We assume that normalized slice emittance is the same
(0.15 mm mrad) as
it was measured in the LCLS injector \cite{20pC} (after compression it was possible to measure only
projected emittance) for 20 pC bunch charge.
Other parameters of the beam and of the undulator are summarized in Table~\ref{tab:sase-u1-par}.

\begin{figure*}[tb]

\includegraphics[width=0.45\textwidth]{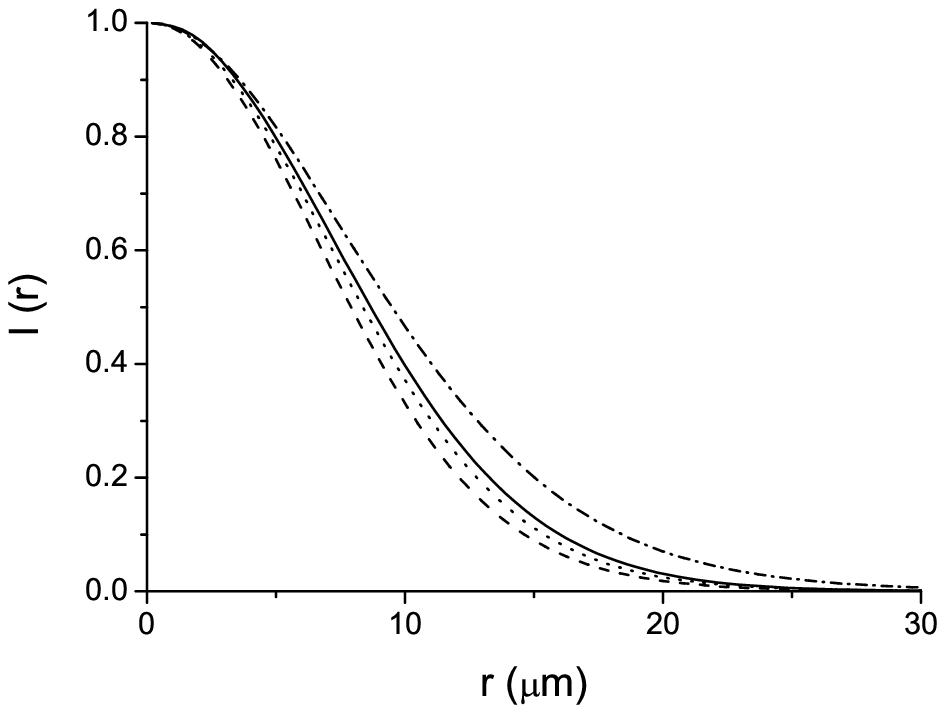}
\includegraphics[width=0.45\textwidth]{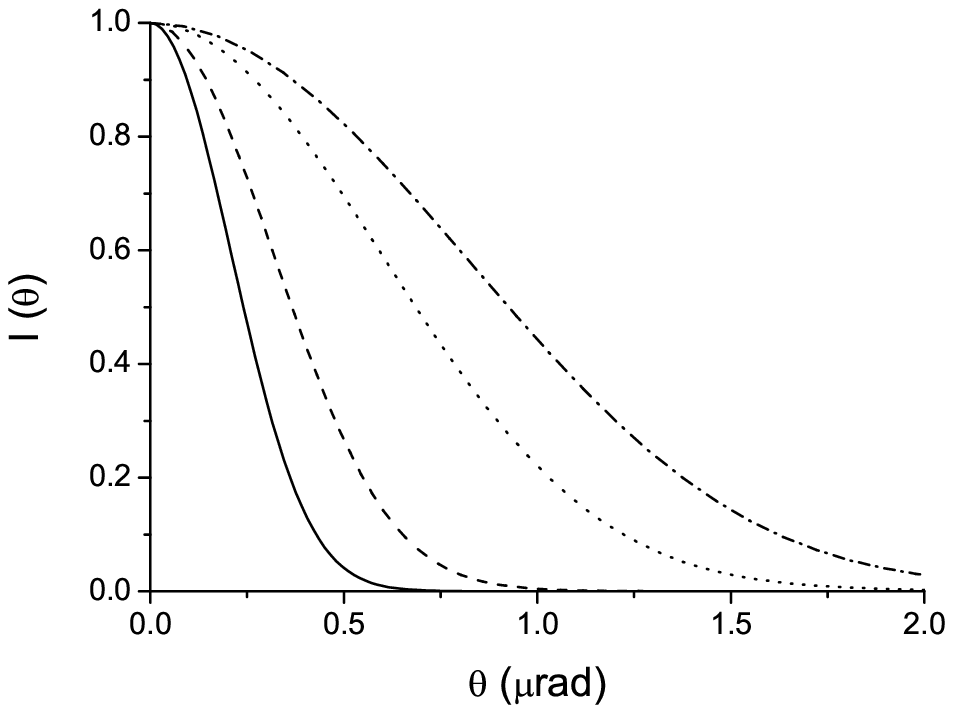}

\caption{\small Intensity distributions at saturation
in the near zone (left plot) and in the far zone (right
plot) for electron beam energy 17.5 GeV and
different photon energies: 20 keV (dash-dot), 30 keV (dot), 60 keV (dash), and 90 keV (solid).}

\label{near-far}
\end{figure*}

We perform numerical simulations of SASE FEL process with the code FAST \cite{fast}.
In Fig.~\ref{sat17-22} we present gain curves for different photon energies and for two different
electron energies. It is seen that a higher electron energy (22 GeV)
is more preferable in the case of higher photon energies. However, the lowest photon energy in this
case is 30 keV (with the smallest undulator gap 6 mm). Beta-function in this calculations was 25 m for
the case 17.5 GeV and 90 keV, and it was 15 m for all other cases.

\begin{figure*}[tb]

\includegraphics[width=0.45\textwidth]{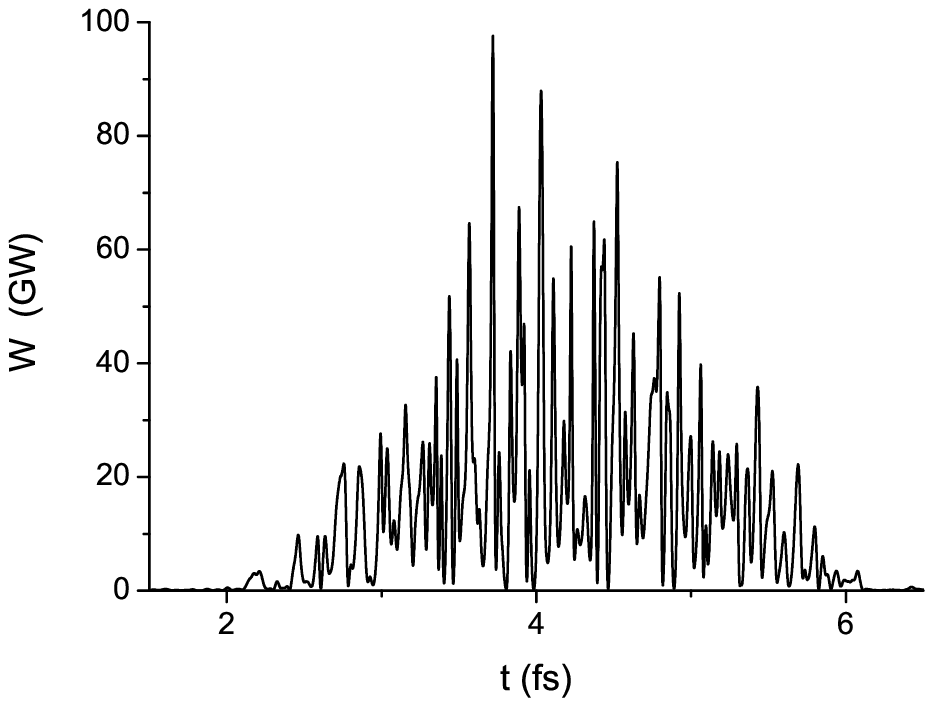}
\includegraphics[width=0.45\textwidth]{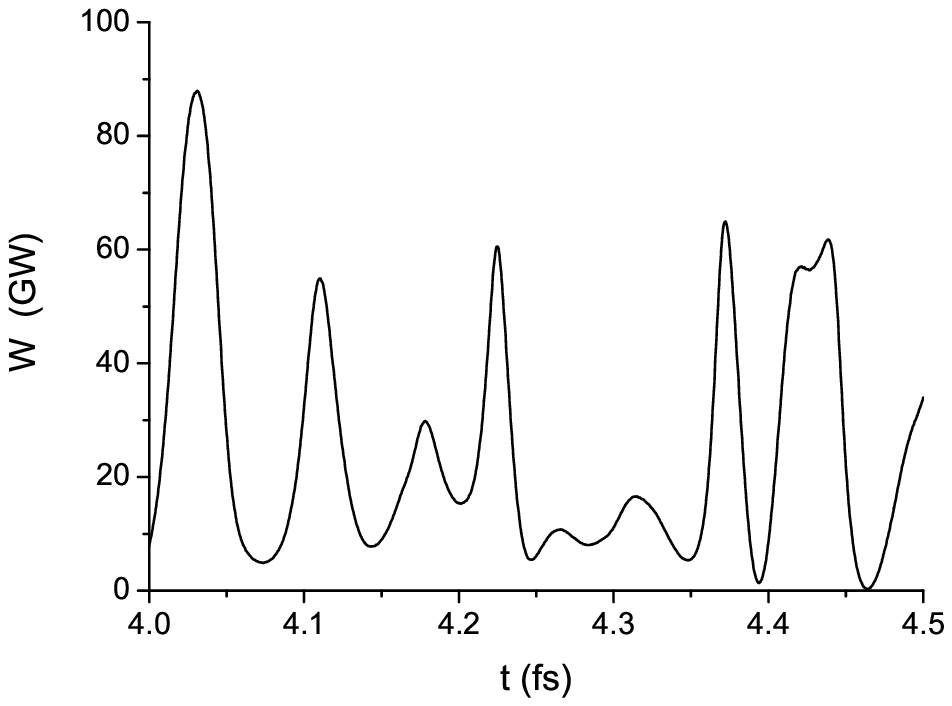}

\caption{\small Temporal structure of an FEL pulse
for electron beam energy 17.5 GeV and photon energy 60 keV at saturation.
Right plot is an enlarged fraction of the left one.}

\label{time-1}
\end{figure*}

\begin{figure*}[tb]

\includegraphics[width=0.45\textwidth]{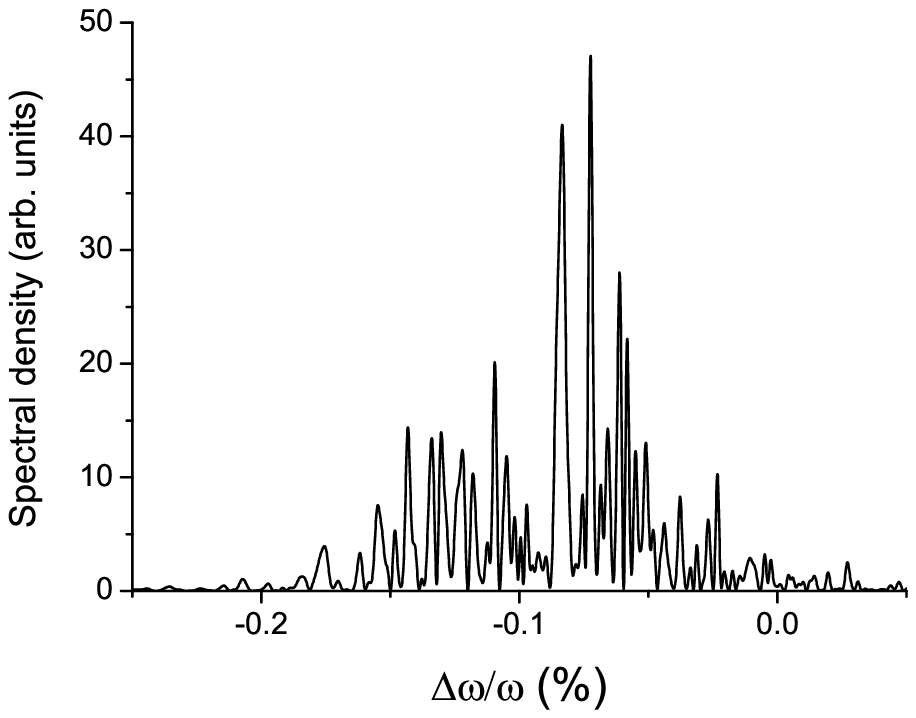}
\includegraphics[width=0.45\textwidth]{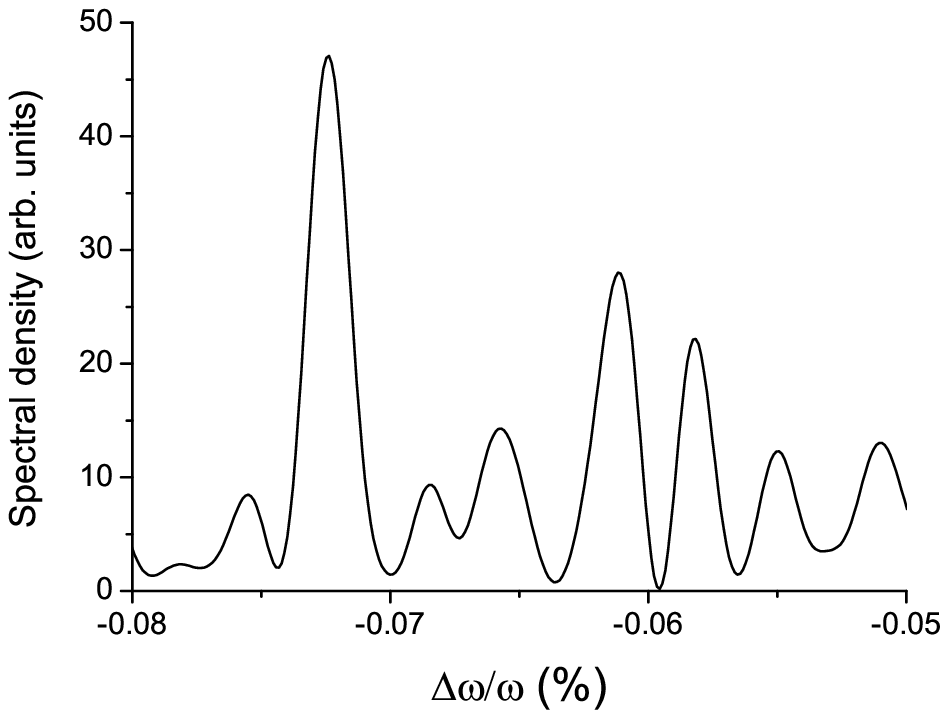}

\caption{\small Spectral structure of an FEL pulse
for electron beam energy 17.5 GeV and photon energy 60 keV at saturation.
Right plot is an enlarged fraction of the left one.}

\label{spec-1}
\end{figure*}

In Fig.~\ref{near-far} we show intensity distributions in the near and far zone for the electron
energy of 17.5 GeV and different photon energies. Corresponding distributions for 22 GeV case are
similar. In Figs.~\ref{time-1}, \ref{spec-1}
we show an example of temporal and spectral intensity distributions
for a specific shot at 60 keV. Peak power is in the range of several tens gigawatts with
pulse duration about 2 fs, and the spectrum width is about 0.1 \%.

\begin{figure*}[tb]

\includegraphics[width=0.7\textwidth]{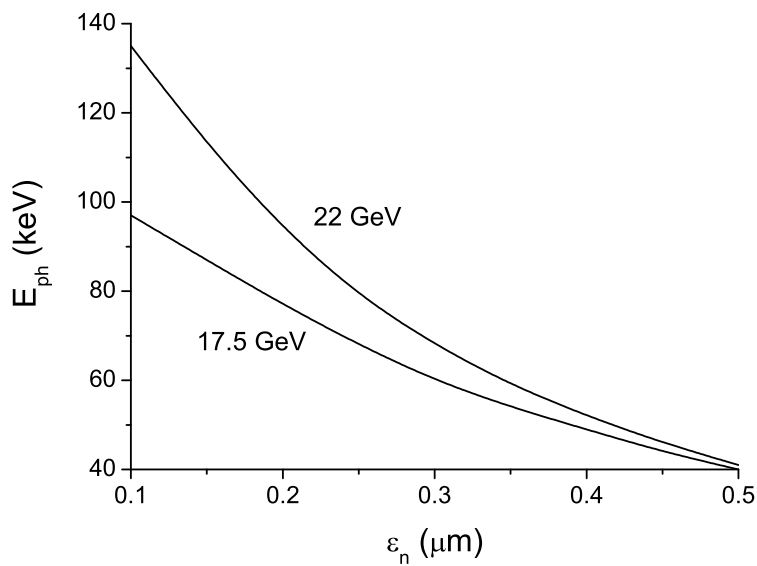}

\caption{\small Photon energy at which SASE FEL saturates within 100 m (net magnetic length)
of SASE-U1 undulator versus
normalized emittance for two electron energies. Other parameters are shown in Table 1.}

\label{sat-emit}
\end{figure*}

We study the dependence on the normalized emittance of the highest photon energy,
that one can reach within the given undulator
length. We do not run numerical simulation code but use universal
design formulas \cite{des-form} instead. The results for two given beam energies are shown in
Fig.~\ref{sat-emit}. One can see, for example, that photon energies
50-70 keV can be reached if normalized slice emittance is 0.3-0.4 mm mrad.
Finally, let us note that influence of quantum effects on the operation
of SASE-U1 undulator are estimated to be
small corrections to the classical description, and can be neglected.

\section{VUV afterburner for SASE-U1}

A simple scheme for SASE afterburner was proposed in \cite{ab-fel,ab-prst}.
Modulations of energy and energy spread,
induced by SASE process on the scale of coherence length, are converted into strong density
modulations on the same scale in a compact dispersion section behind SASE undulator. Then modulated
beam radiates in an undulator or another radiator. The spectrum of modulations is broad, it can
also be easily red-shifted by increasing $R_{56}$ of the dispersion section. With an undulator having
tunable magnetic field, one can therefore cover a wide range of wavelengths. The long-wavelength
pulses are perfectly synchronized with short-wavelength pulses from SASE FEL and can be used in
pump-probe experiments. Optical afterburner for a SASE1 undulator (operating at 1 $\mathrm{\AA}$)
was considered in \cite{ab-fel,ab-prst}. Here we show that a compact afterburner for SASE-U1 can produce tunable
soft x-ray and VUV radiation.

\begin{figure*}[tb]

\includegraphics[width=0.45\textwidth]{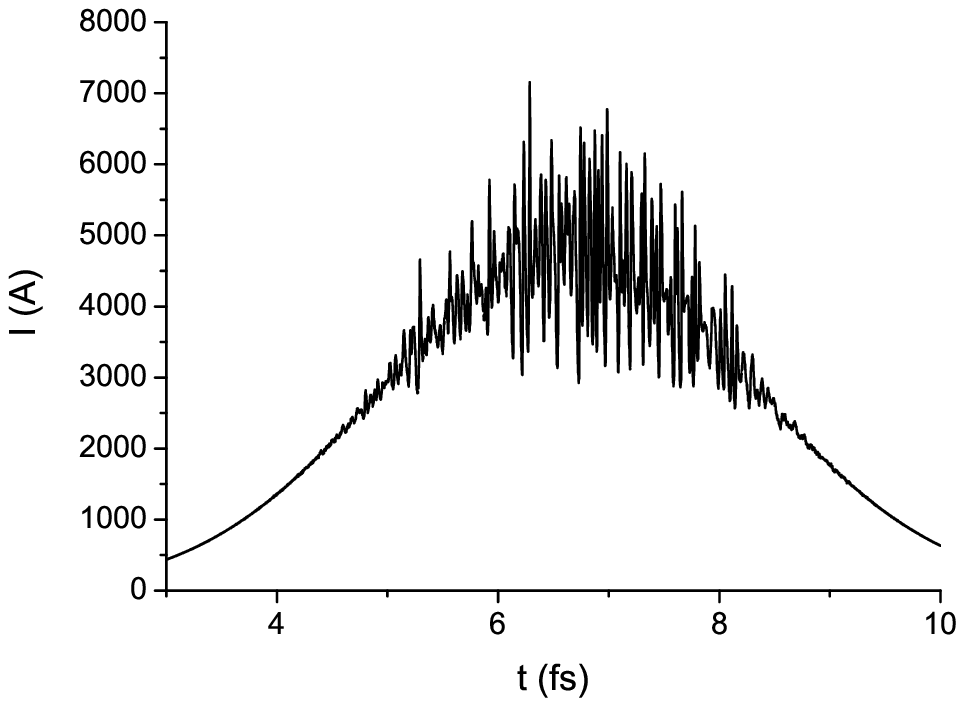}
\includegraphics[width=0.45\textwidth]{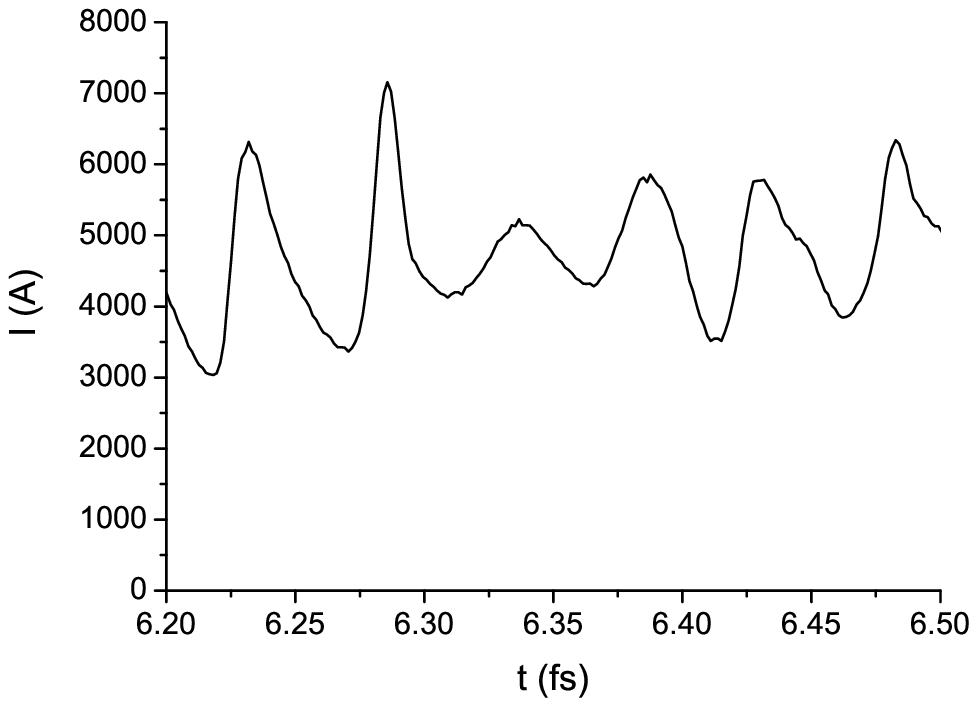}

\caption{\small Current distribution and its enlarged fraction behind a dispersion section with
$R_{56}=$ 2 $\mu$m. SASE-U1 operates at photon energy 60 keV, electron energy is 17.5 GeV.}

\label{ab-time}
\end{figure*}

In Fig.~\ref{ab-time} one can see typical distribution of the beam current behind a chicane
with the $R_{56}=2 \ \mu$m for the case when
SASE-U1 operates at the photon energy of 60 keV with the electron energy of 17.5 GeV.
Typical modulation amplitude is in the range of 20-40 \%.
Spectrum of the beam current is
broadband with the maximum at 15 nm, and there are significant components at 7-10 nm. Maximum of
the spectrum
can be red-shifted by applying larger $R_{56}$ \cite{ab-fel,ab-prst}.
We consider operation of the afterburner in
the current wavelength range of FLASH \cite{njp}, i.e. from 7-10 nm to 40-50 nm.
This can be done by
using, for instance, an undulator with 20 periods, a period length 25 cm, and a total length
$L_{w}=5$ m.
The total length of the afterburner (chicane plus undulator) would be about 10 m. For a given
low-charge beam (20 pC) one can expect (within a central cone)
pulse energies in the range of several hundred nanojouls with a femtosecond duration,
what would correspond to the peak power of several hundred megawatts.
So, the peak power is by an order of magnitude
smaller than that at FLASH \cite{photonics},
and the spectrum width (5 \%) is by an order of magnitude larger.
The peak brilliance is therefore smaller by two orders of magnitude than that of FLASH,
but is  still much larger than the brilliance of other available sources in this wavelength range.
Important is that
these soft x-ray and VUV pulses are perfectly synchronized with hard x-ray pulses from SASE-U1,
and can, therefore, be used in pump-probe experiments with femtosecond resolution. Otherwise
they can be separated and used independently. Let us also note that the minimal achievable wavelength
can be shorter (about 5 nm) for photon energy of 90 keV.

\section{Saturation length in SASE1 and SASE2 undulators versus emittance and wavelength}

\begin{figure*}[b]

\includegraphics[width=0.45\textwidth]{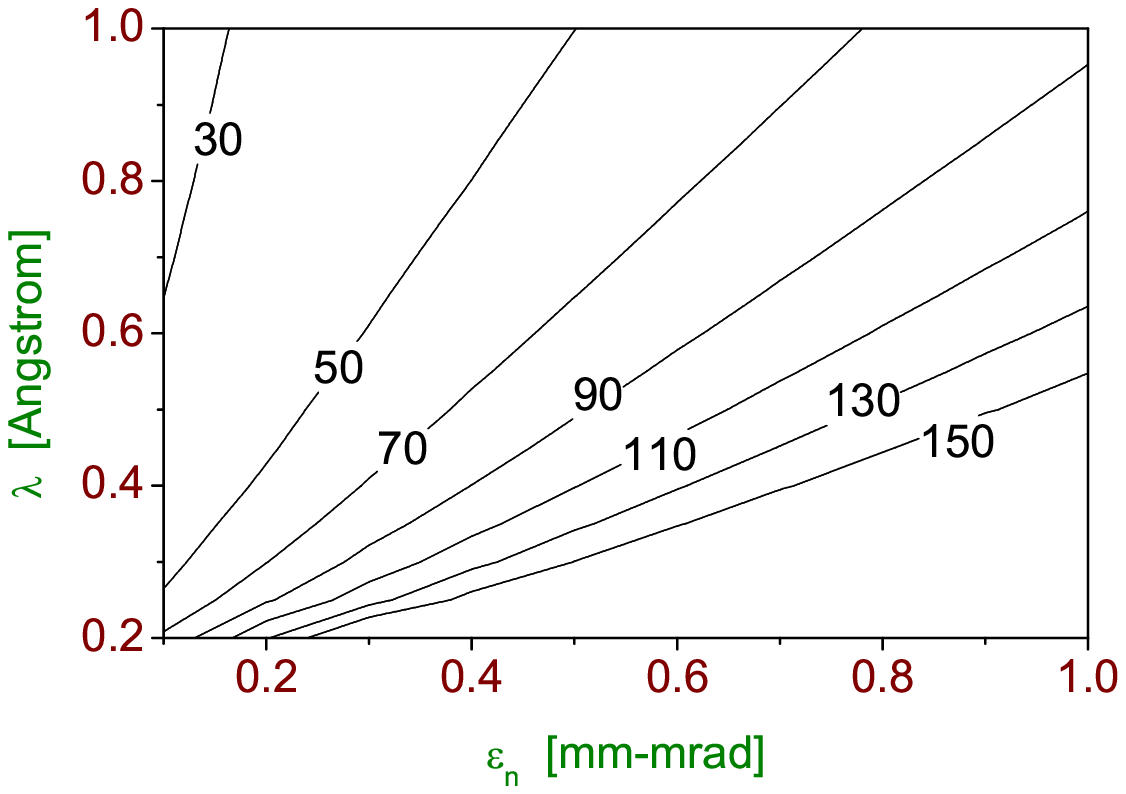}
\includegraphics[width=0.45\textwidth]{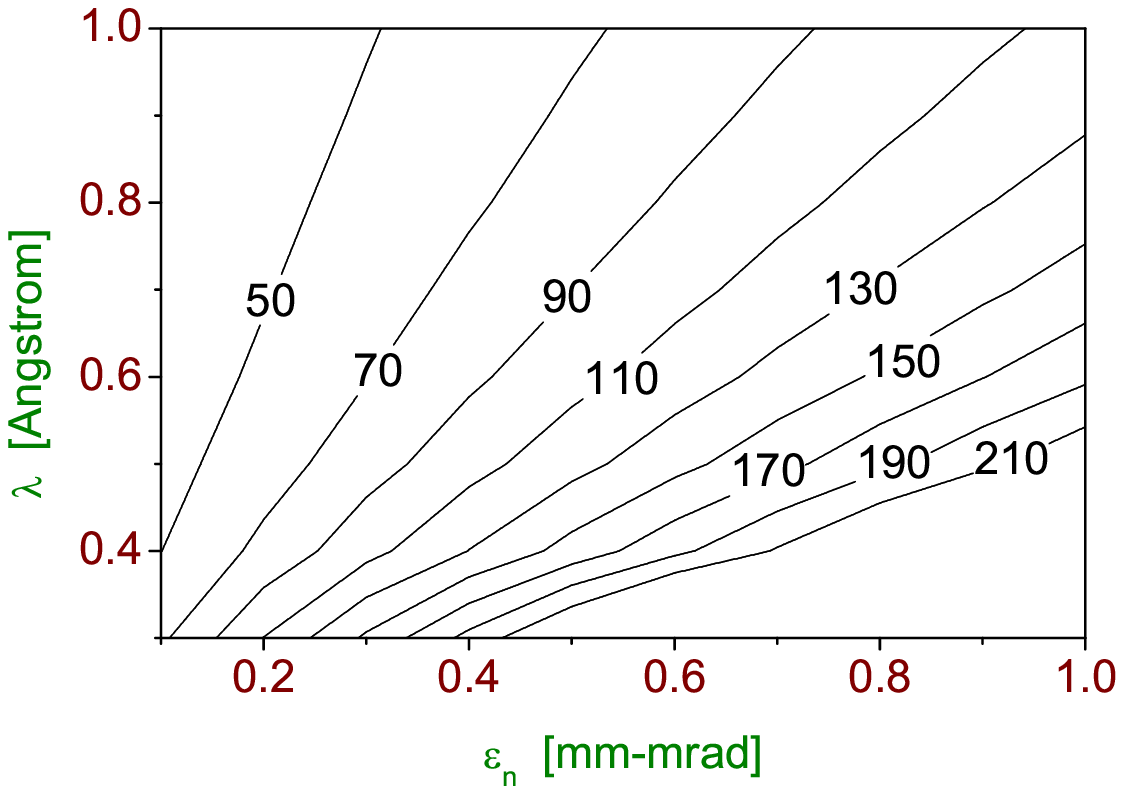}

\caption{\small Contour plots of the saturation length (in meters)
versus normalized emittance and radiation wavelength.
Left and right plots refer to SASE1 and SASE2, respectively. Operating energy of electrons
is equal to 17.5 GeV.
Radiation wavelength is changed by opening the undulator gap.
Betatron function of external focusing is optimized for minimum saturation length.
Technical limit for minimum of the betatron function is set to 10 meters. Other parameters of electron
beam and undulators are given in \cite{euro-xfel-tdr}.}

\label{sase12-sat}
\end{figure*}

The lengths of SASE1 and SASE2 undulators were calculated (with some contingency)
\cite{euro-xfel-tdr}
under the assumption of rather conservative normalized emittance of 1.4 mm mrad, and the shortest
operating wavelength of 1 $\mathrm{\AA}$. If the normalized emittance
is smaller, the
wavelength range can be extended since the gaps of both undulators can be opened
(see Table~\ref{tab:und-param}).
In Fig.~\ref{sase12-sat} we show saturation length (net magnetic
length of the undulators is meant here) in
both undulators versus radiation wavelength and normalized emittance for the nominal beam energy
of 17.5 GeV, peak current of 5 kA, and energy spread of 1 MeV at the entrance of the undulators.
Calculations were done with the help of universal formulas \cite{des-form}.

\begin{table}[b]
\caption{Specification of the SASE undulators at the European XFEL operating
in the extended wavelength range$^{*}$}
\medskip

\begin{tabular}{ l c c c c c c c }
\hline
&
$\lambda _{\mathrm{r}}$ &
$\lambda _{\mathrm{u}}$ &
gap &
$B _{\mathrm{max}}$ &
$K _{\mathrm{rms}}$ &
$L _{\mathrm{w}}$ $^{**}$  \\
& $\mathrm{\AA}$ & mm & mm & T & & m \\
\hline
SASE1 & 0.18-1.0 & 35.6 & 10-28 & 0.2-1 & 0.43-2.4 & 165 \\
SASE2 & 0.24-4.0 & 48 & 10-43 & 0.13-1.3 & 0.42-4.3 & 210 \\
SASE3$^{***}$ & 0.4-16 & 65 & 10-54 & 0.16-1.7 & 0.7-6.7 & 110 \\
SASE-U1 & 0.14-0.6 & 26 & 6-16 & 0.3-1.2 & 0.5-3 & 100 \\
\hline
\end{tabular}

$^*$Wavelength tuning is achieved by extra opening of the undulator gap.
The shortest wavelength is defined by achieving the saturation within total
undulator length at the value of the normalized emittance of 0.15 mm-mrad,
peak current of 5 kA and nominal energy of 17.5 GeV.

$^{**}$ Net magnetic length

$^{***}$ It is assumed that bunches are not disturbed in SASE1. Note also that operation in
hard x-ray range is presently not anticipated.

\label{tab:und-param}

\end{table}

One can see that both undulators hold potential for a significant wavelength reduction.
It is worth noticing, however, that for short wavelengths the undulator K-value gets too small.
Therefore, both SASE1 and SASE2 would profit from operation at higher electron energies
(for which K-value would be larger).

\section{Betatron switcher for a multi-color operation of SASE1 and SASE2 undulators}

\begin{figure*}[tb]

\includegraphics[width=0.7\textwidth]{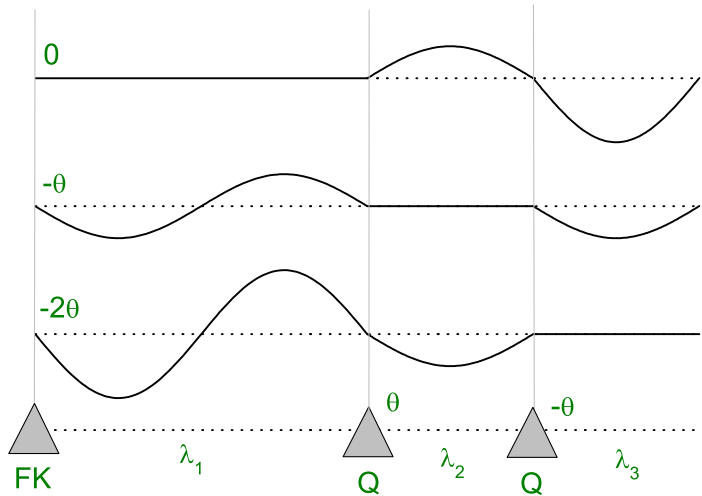}

\caption{\small A schematic illustration of the betatron switcher for
multi-color operation of a SASE undulator.
Here "FK" stands for a fast kicker (giving different kicks to different bunches)
and "Q" for a quadrupole (giving the same static kick to all bunches). Betatron phase advance is
$2\pi$ in the first sub-undulator, and $\pi$ in each of the last two sub-undulators.
Lasing to saturation takes place only on straight sections of beam orbit.}

\label{3-colors}
\end{figure*}

A long undulator with tunable gap (to be specific, in the following we mainly consider SASE2,
although the method can also be applied to
SASE1) may be used for generation of several wavelengths. We propose to install a fast kicker in front
of the undulator (or using the
feedback kicker) and to give different angular kicks
(with a shift between kicks on the order of 10 $\mu$rad or less, depending on requested wavelengths)
to different groups of bunches (a bunch pattern for each
group is defined by users requests). For every group a kick is compensated statically at one location
in the undulator by moving transversely a quadrupole, i.e. by using it as a steerer.
After that location the bunches of this group go straight and lase to saturation in
a part of an undulator (sub-undulator), of which magnetic field is tuned to a desired wavelength
(see Fig.~\ref{3-colors} for illustration).
In other sub-undulators the trajectory of this group strongly deviates from the straight path,
and bunches of this group do not lase. In a
given sub-undulator only one group of bunches lases to saturation, orbits of other groups are
strongly disturbed. So, every group lases in its own sub-undulator, of which magnetic
field is tuned to a requested wavelength.
A length of a sub-undulator is chosen such that a betatron phase advance per
its length is $\pi$ (or multiple of $\pi$) on the one hand, and the length is multiple of
a length of an elementary cell on the other hand.

The elementary cell of SASE2 consists of an
undulator module and a focusing (defocusing) quadrupole (we do not describe here elements that are
not relevant to the operation of the scheme). The length of an elementary cell is $L_{cell}=6.1$ m,
FODO period is equal to $2 L_{cell}$. Beta-function for a given beam energy is defined by the
strength of quadrupoles and can be varied remotely. In the following we assume for simplicity
that the strength is
the same for all quadrupoles in the undulator, so that periodicity is not perturbed.
Optimal beta-function depends on the
wavelength as well as on beam and undulator parameters \cite{des-form}:

\begin{equation}
\beta_{\mathrm{opt}} \simeq 11.2 \left(\frac{I_A}{I} \right)^{1/2} \frac{\epsilon_n^{3/2}
\lambda_{\mathrm{w}}^{1/2}}
{\lambda K A_{JJ}} \ (1+8\delta)^{-1/3}
\label{beta}
\end{equation}

\noindent Here $I$ is the beam current, $I_A = 17$ kA is Alfven current, $\epsilon_n$ is the
normalized emittance, $\lambda$ is resonant wavelength, $\delta$ is a parameter depending on
energy spread \cite{des-form} (usually a small correction),
$\lambda_{\mathrm{w}}$ is the undulator period,
$K$ is the rms undulator parameter, $A_{JJ} = 1$ for a helical
undulator and $A_{JJ} = J_0(K^2/2(1+K^2))-J_1(K^2/2(1+K^2))$ for a planar undulator,
$J_0$ and $J_1$ are the Bessel functions of the first kind. The expression (\ref{beta}) was obtained
in \cite{des-form} under an assumption of small betatron phase advance per FODO period,
$2 L_{cell} \ll \beta$.
If this condition is not satisfied, this expression is still a good first guess for an average
beta-function. Also note that this condition is not necessary for operation of the proposed scheme.

As one can see from (\ref{beta}), optimal beta function is inversely proportional to the wavelength. We
choose an optimal beta-function for a shortest wavelength from a requested set because FEL saturation
length is the largest for the shortest wavelength. Note, however that deviations at 10-20 \% level are
tolerable (since a function changes slowly near an optimum)\footnote{If the undulator length allows,
one can even use $\beta$ that deviates significantly from the optimum for the shortest wavelength.},
so that one has some freedom to adjust $\beta$.
Then we define a length of a shortest possible sub-undulator as\footnote{For simplicity
we assume here that a kick is localized just in front of the
undulator. In a general case one might think of a (tunable) phase advance between the kicker and
the undulator, which is multiple of a phase advance per cell. In that case the length of the
first sub-undulator is reduced, what might be tolerable for the longest requested wavelength}

\begin{displaymath}
\int\limits_0^{L_{sub}^0} \frac{d z}{\beta} = \pi
\end{displaymath}

\begin{displaymath}
L_{sub}^0 = n L_{cell}
\end{displaymath}

\noindent Under these conditions
the integrated kick from upstream quadrupoles and a current quadrupole
(all located at zero crossings of electron orbit)
can compensate
exactly a kick from fast kicker for a given group of bunches. Then these bunches go straight and
lase in a given sub-undulator.
A length of a sub-undulator, depending on a wavelength, can be a multiple of the elementary
sub-undulator length, $L_{sub} = m L_{sub}^0$.

Let us consider a numerical example, using the formulas of Ref. \cite{des-form}.
Consider an electron beam with the following parameters:
$I=$ 5 kA, $\epsilon_n = 0.4$ mm mrad, energy spread is 1 MeV at the entrance of SASE2, beam energy
is 17.5 GeV. The length of SASE2 undulator is 256 m (42 cells).
Imagine that users request three different wavelengths (see Fig.~\ref{3-colors} for illustration):
0.5 $\mathrm{\AA}$, 1.6 $\mathrm{\AA}$, and 2.3 $\mathrm{\AA}$.
Optimal beta-function for 0.5 $\mathrm{\AA}$ is about 20 m, we adjust it such that the above mentioned
conditions are met. As a result, we choose an elementary sub-undulator to be equal to 10 cells (61 m).
In order to have FEL saturation at 0.5 $\mathrm{\AA}$, we use two elementary sub-undulators, i.e. the
magnetic field of first 20 cells is tuned for this wavelength ($K=1.2$). The next 10
cells are tuned to 1.7 $\mathrm{\AA}$ ($K=2.7$), and the 10 cells after that operate at
2.3 $\mathrm{\AA}$ ($K=3.2$). Although beta-function is significantly larger that the
optimal ones in the last two cases, the wavelengths are long enough for saturation within given
sub-undulators. In this case the group of bunches, lasing at 0.5 $\mathrm{\AA}$, is undisturbed
by a kicker, the second group (1.7 $\mathrm{\AA}$) gets a kick of -10 $\mu$rad, and the third one
-20 $\mu$rad.\footnote{Kicks to the one direction are considered here for simplicity.
Note that in this specific example the symmetric kicks of $\pm 10 \ \mu$rad are also possible.}
After 20 cells (full period of betatron oscillations) the
quadrupole gives the kick of 10 $\mu$rad to all bunches, so that now the second group goes straight,
but the first and the third have $\pm 10 \ \mu$rad and do not lase. After 30 cells the quadrupole
compensates the kick for the third group, then the first and the second ones have -20 and -10 $\mu$rad,
respectively. The last two cells are not used. Also note that in the case when unspoiled bunches
for SASE-U1 are needed, in the considered case this would be the fourth group that gets a kick of
-30 $\mu$rad, and the kick is compensated, for instance, after 40 cells. Thus, in SASE-U1 only
this unspoiled group lases but the other groups do not.
A number of possible colors is mainly defined by the shortest wavelength.
It can be increased to 4-5 in SASE2 if the shortest  wavelength is about 1.5 $\mathrm{\AA}$ or larger.
Alternatively, it can be increased for a smaller emittance (see Fig.~\ref{sase12-sat}).

Let us now consider an example for SASE1 undulator for the same parameters of the electron beam as
in the example with SASE2. SASE1 consists of 33 cells (total length is 201 m).
At the nominal operating energy of 17.5 GeV the longest possible wavelength (with closed undulator
gap of 10 mm) is 1 $\mathrm{\AA}$. Operation at shorter wavelengths is possible
(see Fig.~\ref{sase12-sat}) by opening the gap. In our example the first part of the undulator
(16 cells with net magnetic length 80 m and total length about 98 m) is tuned to a resonance with
0.5 $\mathrm{\AA}$, beta-function is close to 16 m (phase advance per 16 cells is $2\pi$). The
second part of the undulator can be used for generation of another x-ray beam
with any wavelength between 0.5 $\mathrm{\AA}$ and 1 $\mathrm{\AA}$. The number of colors can be
increased for a smaller emittance.

It is worth to note that the proposed scheme is very simple and robust.
A fast kicker is compact (about 1 m long) and not expensive,
one can use the same type as that used in a separation system upstream of SASE undulators.
High accuracy of kicks is not required, a per cent level is tolerable.

Distribution of x-ray beams with different wavelengths can be based on multilayer movable
mirrors \cite{mult-user}. A disadvantage of such a scheme is that
an entire macropulse goes to a single user, so that, for instance, for 5-color operation of
the undulator, each
user station gets macropulses with the repetition rate of 2 Hz instead of possible 10 Hz. In addition,
multilayer mirrors must be exchanged when a given user station is supposed to
run with a new wavelength. Here we would like to attract an attention to another option of distribution
of photon beams. Namely, one can make use of a recently developed x-ray prism (see \cite{tur}
and references therein), which is made of high-quality diamond and can operate in the range from
2-4 keV to 100 keV. A resolution of $10^3 - 10^4$ is claimed \cite{tur} so that, in principle,
even very near colors can be separated.
A long transport line of the European XFEL should be sufficient for spacial
separation of dispersed x-ray beams of different colors - and, of course, an actual geometry would
define how near the colors could be. In case of using such kind of prism every user gets a
required pulse
pattern with the repetition rate of 10 Hz. This option, however, requires futher studies.

Using the principle of the betatron switcher, we can also propose a scheme for pump-probe
experiments. A long flat-top laser pulse (20-25 ps) in the XFEL injector is formed from many short
pulses (about 2 ps). One can program the laser operation such that, for instance, only two of such
pulses with variable separation between them are produced. Two low-charge electron bunches are then
produced in photoinjector, and compressed with the help of linearized bunch compression system.
The distance between bunches is compressed proportionally, so that finally one can vary the
separation on the scale of tens (or hundreds) of femtoseconds. In front of an undulator one of
these two bunches gets a small angular kick from a transverse deflecting cavity \cite{tds}, while
the other one is not kicked. The kick is compensated statically at some position in
the undulator as described above for the scheme, based on fast kicker (so the deflecting
cavity in this scheme is just an ultrafast kicker). Two bunches produce then two different
colors. In principle, the scheme can be generalized for the case of several bunches as described above.
Thorough analysis of this scheme will be presented elsewhere.

\section{Discussion on extension of wavelength range of the European XFEL facility}

In this paper we have considered only few of many possibilities that
can be adopted by the European XFEL in view
of recent achievements of LCLS team and rapidly developing physics and technology of
high-brightness electron beams. Here we would like to discuss how the facility
can cover with powerful, coherent
radiation a part of the electromagnetic spectrum from far infrared to gamma-rays without
a significant modification of the facility layout and without
using external sources of radiation.

The electron bunches are longitudinally
compressed down to ten or even one micron size, they can radiate coherently at wavelengths that are
longer than their size (using parasitic or dedicated radiators of different types), so that
practically the entire infrared range can be covered.
Operation of SASE afterburners \cite{ab-fel,ab-prst} for different SASE undulators is possible
from a few microns (SASE1, SASE2, and SASE3) down to 5-10 nm (SASE-U1).
In this paper we have proposed to combine U1 and U2 undulators in one beamline (SASE-U1).
Then U2 part (see Fig.~\ref{xfel}) can be used
for installation of the proposed SASE4 undulator \cite{sase4} that can deliver sub-terawatt level of
peak power in the range 1.6-6.4 nm.
Other SASE undulators (SASE1, SASE2, SASE3 and SASE-U1) would cover the rest of x-ray range
with coherent harmonics from SASE-U1 penetrating into gamma-ray range (beyond 100 keV).

\end{document}